\documentclass[12pt]{article}
\usepackage{latexsym, epsfig, graphics}
\newcommand{\be}{\begin{equation}}
\newcommand{\ee}{\end{equation}}
\newcommand{\bq}{\begin{eqnarray}}
\newcommand{\eq}{\end{eqnarray}}
\newcommand{\cue}{{\bf q}}
\newcommand{\dbsm}{\sum_{\sigma_{1}<\sigma}\sum_{\sigma<\sigma_{2}}}
\newcommand{\sqsm}{\sqrt{(\sigma-\sigma_{1})(\sigma_{2}-\sigma_{1})
(\sigma_{2}-\sigma)}}
\newcommand{\wss}{W(\sigma_{1},\sigma_{2})}
\newcommand{\ssm}{\frac{\sigma_{2}-\sigma_{1}}{(\sigma_{2}-\sigma)
(\sigma -\sigma_{1})}}
\newcommand{\bfv}{{\bf v}}
\begin{document}
\begin{titlepage}
\today          \hfill 
\begin{center}

\vskip .5in

{\large \bf Field Theory On The World Sheet: Improvements And
Generalizations}
\footnote{Notice:* This manuscript has been authored by Korkut Bardakci
under Contract No.DE-AC02-05CH11231 with the U.S. Department of Energy.
The United State Goverment retains and the publisher, by accepting the
article for publication, acknowledges that the United States Goverment
retains a non-exclusive, paid-up, irrecovable, world-wide license to
publish or reproduce the published of this manuscript, or allow others
to do so, for United States Goverment purposes.}
\vskip .50in


\vskip .5in
Korkut Bardakci \footnote{Email: kbardakci@lbl.gov}

{\em Department of Physics\\
University of California at Berkeley\\
   and\\
 Theoretical Physics Group\\
    Lawrence Berkeley National Laboratory\\
      University of California\\
    Berkeley, California 94720}
\end{center}

\vskip .5in

\begin{abstract}

This article is the continuation of a project of investigating planar
$\phi^{3}$ model in various dimensions. The idea is to 
reformulate them on the world sheet, and then to apply the classical
(meanfield) approximation, with two goals: To show that the ground
state of the model is a solitonic configuration on the world sheet, and
 the quantum fluctuations around the soliton lead to the formation
of a transverse string. After a review of some of the earlier work,
we introduce and discuss several generalizations and
 new results. In 1+2 dimensions,
a rigorous upper bound on the solitonic energy is established.
A $\phi^{4}$ interaction is added to stabilize the original
$\phi^{3}$ model. In 1+3 and 1+5 dimensions, an improved treatment
of the ultraviolet divergences is given. And significantly, we show that
our approximation scheme can be imbedded into a systematic
strong coupling expansion. Finally, the spectrum of
quantum  fluctuations around the soliton confirms earlier results:
In 1+2 and 1+3 dimensions, a transverse string is formed on the world
sheet.

\end{abstract}
\end{titlepage}

\newpage
\renewcommand{\thepage}{\arabic{page}}
\setcounter{page}{1}
\noindent{\bf 1. Introduction}
\vskip 9pt

The present article is in part a review and summary of an ongoing
project to analyze field theory from a world sheet perspective.
The original idea was to sum the planar graphs of the $\phi^{3}$ field
theory in both $3+1$ and $5+1$ dimensions, starting with the world sheet
picture developed in [1], which in turn was based on the pioneering
work of 't Hooft [2]. This project went through  many
phases, and somewhat varied versions  have appeared. In this paper,
we go back to the references [3] and [4] as our starting point, and our goal
will be to present a fresh approach 
to resolve some of the problems left open in these references.

In section 2, we review of the world sheet picture of the planar
 graphs of the $\phi^{3}$ field theory that is our starting point,
 and in section 3, we describe
the field theory on the world sheet, developed in [3], which reproduces
these graphs. This theory is formulated in terms of a complex scalar
field and a two component fermionic field; a central role is played by the
field $\rho$ (eq.(4)), a composite of the fermions, which
roughly measures the density of graphs on the world sheet.  $\rho_{0}$,
the ground state expectation value of $\rho$, turns out to be an
 order parameter that distinguishes between two phases. A non-zero value for 
 $\rho_{0}$ corresponds to a phase in which
the world sheet densely covered with graphs, and an
 important question is whether this phase 
 has lower energy than the trivial phase (empty
world sheet) $\rho_{0}=0$.
  An old idea
 that a densely covered
world sheet would naturally have a string description motivated
 some of the very early work on this subject [5, 6].
 To find such a
string picture  has been a
goal of the present, as well as of the earlier work.

To find out which phase is energetically favored, a variational
calculation was carried out in [3], and $\rho_{0}\neq 0$ was found
to have lower energy than the trivial ground state. In [4], the same
problem was tackled using a different approach: A static 
classical solution on the world sheet, with again $\rho_{0}\neq 0$
and with energy lower than the trivial ground state, is constructed.
These two approaches, although seemingly different, are actually closely
 related.

The computations discussed above suffer from two kinds of
divergences: One of them is the standard field theoretic ultraviolent
divergence which we will address later on. The second one is an 
 infrared divegence due to the choice of the light cone
coordinates. This infrared problem is temporarily
 circumvented by the discretizing the $\sigma$ coordinate on the world sheet
in steps of $a$, which amounts to the compactification of the the light cone
coordinate $x^{-}$. One place where the infrared divergence manifests
itself is in the ground state energy in the non-trivial phase; it is
negative and proportional to $1/a^{2}$. Two questions present
themselves: Is this a real effect or merely a spurious infrared
divergence? And if it is a real effect, is it somehow connected with
classical instability of the $\phi^{3}$ theory?

In section 4, we answer the first question: It is a real physical
effect. The model we study is $\phi^{3}$ in $D=1$ (1+2) dimensions.
We study this lower dimensional field theory because it is free of
ultraviolet divergences, and therefore
  we can focus on the infrared problem
without being distracted by  ultraviolet divergences. We  use the
variational method to estimate the ground state energy and find that
the phase $\rho_{0}\neq 0$ is energetically preferred
 and the corresponding ground state energy is negative and 
proportional to $1/a^{2}$.
 This approach has the
advantage of providing a rigorous upper bound for the energy; therefore,
the divergence in the $a\rightarrow 0$ limit is not a spurious effect
  resulting from the approximation scheme used.
At the end of the section, we give a explanation of the negative sign of the
energy based on statistical mechanics, and we also observe that this
 divergence
can be cancelled by adding a counter term proportional to the area of
the world sheet.

In section 5, still keeping $D=1$, we add a $\phi^{4}$ term with positive
coefficient to the interaction. Using the same variational method, we find the
same $- 1/a^{2}$ behaviour in the ground state energy. This answers the 
second question posed earlier: Since we now have at least classically
stable theory, the behaviour of the ground state energy is not
related to the instability of $\phi^{3}$.   

Next, in section 6, we consider $\phi^{3}$ in $D=2$ (1+3) dimensions. 
Although we could use  the variational method, in this case, it is more
convenient to search for a classical static solution (soliton)
on the world sheet. Such a solution was already found in [4], and
the corresponding ground state energy had a logarithmic
ultraviolet divergence. This the standard perturbation result, and 
the renormalization prescription is to eliminate it by a mass
counter term. Unfortunately, the structure of the divergent term
is different from the structure of the mass term, so the
standard cancellation does not work. This problem can be traced back
to the construction of the classical solution. In particular, $\rho$ is
treated as a classical continuous variable, whereas it is in reality
it takes on only the discrete values 0 and 1. Consequently, certain
overlap identities (eqs.(47) and (48)), which are crucial for the
correct structure of the ultraviolet divergence, are violated.
We overcome this problem by using the exact overlap identities in the
initial stages of the calculation, and introducing the classical
approximation only after taking care of the divergent term by
mass renormalization. After this, things work pretty much as in the
case $D=1$: The phase $\rho_{0}\neq 0$ is energetically preferred
 and the corresponding ground state energy is negative and 
proportional to $1/a^{2}$.

There are further complications related to renormalization in
the case of $1+5$ dimensions ($D=4$), which are addressed in section 7.
The mass term is now quadratically divergent, and there is also
a logarithmic coupling constant divergence. The classical solution is
 constructed using exact overlap relations (eqs.(47) and (48)) in the
initial stages of the computation, same as in the previous section.
Their structure is then the same as in perturbation theory, and they can
 readily be renormalized. The final result is, however, different
from the cases $D=1$ and $D=2$: The ground state energy is given
solely by a renormalized mass term. If the square of the mass is positive,
the trivial ground state $\rho_{0}=0$ is energetically favored.
The phase of interest, $\rho_{0}\neq 0$, is energetically favored
only if the renormalized mass is tachyonic. Whether this is
physically sensible is an open question.

Section 8 deals with the quadratic quantum fluctuations around the
classical solutions. This section is largely based on references [4,7]
and is included here for the sake of completeness. In the cases of
$D=1$ and $D+2$, the calculations are straightforward, and the
results can be summarized as follows: The spectrum consists of
two components; the heavy sector and the light sector. The energies
of the heavy states go to infinity as $a\rightarrow 0$.
Since the radius of compactification of the coordinate
$x^{-}$ is $1/a$, it is natural to identify them with Kaluza-Klein
states whose masses go to infinity in the decompactification limit.
 In contrast,
the light sector, represented by a transverse string
with a finite slope, remains finite. Thus, the original motivation
for putting field theory on the world sheet, the formation of a string,
 is realized
at least in lower dimensions. In contrast to the lower
dimensional cases, string formation at $D=4$ remains problematic,
since it requires  the renormalized mass to be tachyonic.

The existence of the light sector is the direct consequence of
translation invariance in the variable $\cue$ (eq.(17)), which
forbids a heavy mass term. Although the original action for the
model (eq.(16)) was non-local in the coordinate $\sigma$, the 
string action is local in this coordinate. This localization is a
consequence of $\rho_{0}$ being non-zero: Terms seperated by a finite
distance in $\sigma$ are suppressed [7]. 

 The expansion around the classical configuration
described above is not a
systematic expansion, since we have not identified an expansion parameter.
In section 9, at least in the case of $D+1$, we remedy this defect;
The  parameter 
$$
e^{2}= g^{- 4/3},
$$
where $g$ is the coupling constant
(eq.(82)), serves as an expansion parameter.
 It therefore  a strong coupling expansion. We show that,
by suitably scaling the parameters of the model, the mass of the soliton
 also scales in expected fashion (eq.(83)). 

Since part of this article is a review of  references [3] and [4],
we would like to draw attention to  the new material included 
here which go beyond those references. The addition of a $\phi^{4}$
term as an interaction is such a new feature. It
  stabilizes the original $\phi^{3}$ model, and it can serve as a
first step towards more realistic models, which usually have such a
term. Another new feature is the treatment of the ultraviolet 
divergences in mass in $1+3$ and $1+5$ dimensions: These divergences can now
be cancelled by introducing mass counter terms in the original action.
Finally, we feel that the inroduction of a systematic expansion
around the classical solitonic configuration is an important new
development; it opens up the prospect of computing terms higher order than
quadratic. It may also make it possible to investigate questions
such as Lorentz invariance in a systematic fashion.

\vskip 9pt
\noindent{\bf 2. The World Sheet Picture}
\vskip 9pt

The planar graphs of $\phi^{3}$ can be represented [4] on a world sheet
parameterized by the light cone coordinates $\tau=x^{+}$ and
$\sigma=p^{+}$ as a collection of horizontal solid lines (Fig.1), where
the n'th line carries a D dimensional transverse momentum $\cue_{n}$.
\begin{figure}[t]
\centerline{\epsfig{file=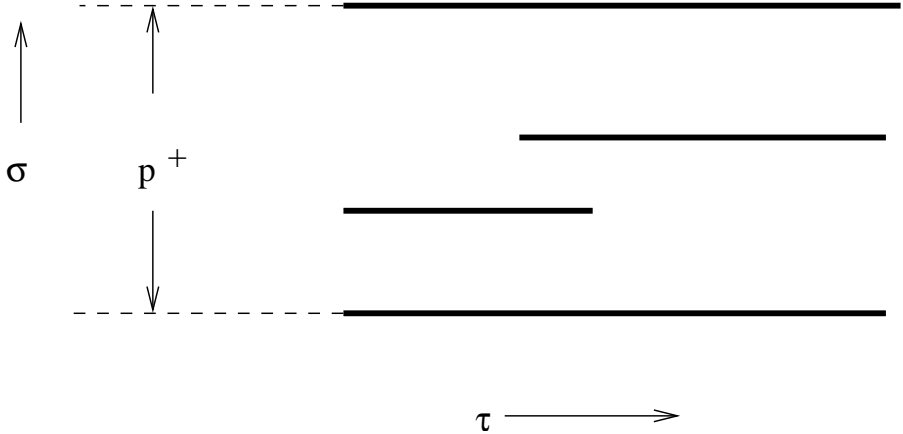, width=12cm}}
\caption{A Typical Graph}
\end{figure}
Two adjacent solid lines labeled by n and n+1 correspond to the light
cone propagator
\be
\Delta({\bf p}_{n})=\frac{\theta(\tau)}{2 p^{+}}\,\exp\left(
-i \tau\, \frac{{\bf p}_{n}^{2}+ m^{2}}{2 p^{+}}\right),
\ee
where ${\bf p}_{n}= \cue_{n}-\cue_{n+1}$ is the momentum flowing through
the propagator. A factor of the coupling constant g is inserted
 at the beginning and at the end of each line, where the interaction
takes place. Ultimately, one has to integrate over all possible
locations and lengths of the solid lines, as well as over the
momenta they carry.

The propagator (1) is singular at $p^{+}=0$. It is well known that 
this is a spurious singularity peculiar to the light cone picture.
To avoid this singularity, and as well as other technical reasons,
it is convenient to temporarily
discretize the $\sigma$ coordinate in steps of length $a$.
This amounts to compactifying the light cone coordinate
$x^{-}$ at radius $R=1/a$. This sort of compactification has been
extensively used both in field theory [8] and in string theory [9]. 
A useful way of visualizing the discretized world sheet is
pictured in Fig.2. The boundaries of the propagators are marked by
solid lines as before, and the bulk is filled by dotted lines spaced
at a distance $a$. For the time being, we will keep $a$ finite, and
later, we will discuss the limit $a\rightarrow 0$.
For convenience, the $\sigma$ is compactified by imposing periodic
boundary conditions at $\sigma=0$ and $\sigma=p^{+}$. In contrast, the
boundary conditions at $\tau=\pm \infty$ are left arbitrary.
\begin{figure}[t]
\centerline{\epsfig{file=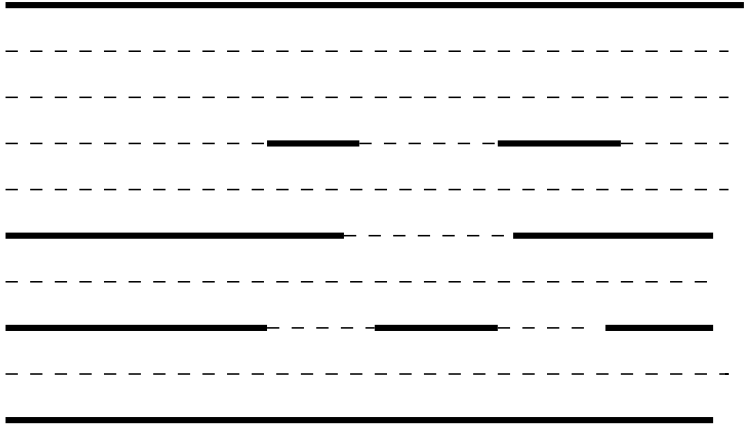, width=10cm}}
\caption{Solid And Dotted Lines}
\end{figure}

\vskip 9pt

\noindent{\bf 3. The World Sheet Field Theory}

\vskip 9pt

It was shown in [3] that the light cone graphs described above are
reproduced by a world sheet field theory, which we now briefly review.
We introduce the complex scalar field $\phi(\sigma,\tau,\cue)$ and
its conjugate $\phi^{\dagger}$, which at time $\tau$
 annihilate (create) a solid line with coordinate $\sigma$ carrying
momentum $\cue$. They satisfy the usual commutation relations
\be
[\phi(\sigma,\tau,\cue),\phi^{\dagger}(\sigma',\tau,\cue')]=
\delta_{\sigma,\sigma'}\,\delta(\cue-\cue').
\ee
The vacuum, annihilated by the $\phi$'s, represents the empty world sheet.

In addition, we introduce a two component fermion field $\psi_{i}(
\sigma,\tau)$, $i=1,2$, and its adjoint $\bar{\psi}_{i}$, which
satisfy the standard anticommutation relations. The fermion with
$i=1$ is associated with the dotted lines and $i=2$ with the solid
lines. The fermions are needed to avoid unwanted configurations
on the world sheet. For example, multiple solid lines generated by
the repeated application of $\phi^{\dagger}$ at the same $\sigma$
would lead to overcounting of the graphs. These redundant states can
be eliminated by imposing the constraint
\be
\int d\cue\, \phi^{\dagger}(\sigma,\tau,\cue)\phi(\sigma,\tau,\cue)
=\rho(\sigma,\tau),
\ee
where
\be
\rho=\bar{\psi}_{2}\psi_{2},
\ee
which is equal to one on solid lines and zero on dotted lines. This
constraint ensures that there is at most one solid line at each
site.

Fermions are also needed to avoid another set of unwanted configurations.
Propagators are assigned only to adjacent solid lines and not to
non-adjacent ones. To enforce this condition, it is convanient to
define, 
\be
\mathcal{E}(\sigma_{i},\sigma_{j})=\prod_{k=i+1}^{k=j-1}\left(
1-\rho(\sigma_{k})\right),
\ee
for $\sigma_{j}>\sigma_{i}$, and zero for $\sigma_{j}<\sigma_{i}$.
The crucial property of this function is that it acts as a projection: 
It is equal to one when the two lines at $\sigma_{i}$ and $\sigma_{j}$
are seperated only by the dotted lines; otherwise, it is zero. With the
help of $\mathcal{E}$, the free Hamiltonian can be written as
\bq
H_{0}&=&\frac{1}{2}
\sum_{\sigma,\sigma'}\int d\cue \int d\cue'\,\frac{\mathcal
{E}(\sigma,\sigma')}{\sigma'-\sigma} \left((\cue-\cue')^{2}+ m^{2}
\right)\nonumber\\
&\times& \phi^{\dagger}(\sigma,\cue) \phi(\sigma,\cue)
 \phi^{\dagger}(\sigma',\cue') \phi(\sigma',\cue')\nonumber\\
&+&\sum_{\sigma} \lambda(\sigma)\left(\int d\cue\,
 \phi^{\dagger}(\sigma,\cue) \phi(\sigma,\cue) -\rho(\sigma)\right),
\eq
where $\lambda$ is a lagrange multiplier enforcing the constraint (3).
The evolution operator $\exp(-i \tau H_{0})$, applied to states,
generates a collection of free propagators, without, however, the
prefactor $1/(2 p^{+})$.

One can also think of 
the lagrange multiplier $\lambda(\sigma, \tau)$ as an abelian
gauge field on the world sheet. The corresponding gauge
transformations are [4]
\bq
\psi &\rightarrow& \exp\left(- \frac{i}{2} \alpha\, \sigma_{3}\right)\,
\psi, \,\,\, \bar{\psi} \rightarrow \bar{\psi}\,\exp\left(\frac{i}{2}
\alpha\, \sigma_{3}\right), \nonumber\\
\phi &\rightarrow& \exp( -i \alpha)\,\phi,\,\,\,
 \phi^{\dagger} \rightarrow
\exp(i \alpha)\,\phi^{\dagger},\nonumber\\
\lambda &\rightarrow& \lambda - \partial_{\tau} \alpha.
\eq
In what follows, we will usually gauge fix by setting
$$
\rho_{+}=\rho_{-},
$$
where
\be
\rho_{+}=\bar{\psi}_{1} \psi_{2},\,\,\,\rho_{-}=\bar{\psi}_{2}
\psi_{1}.
\ee

Using the constraint (3), the free hamiltonian can be written in a
form more convenient for later application:
\bq
H_{0}&=&\frac{1}{2}\sum_{\sigma,\sigma'}G(\sigma,\sigma')\Bigg(
\frac{1}{2} m^{2}\,\rho(\sigma) \rho(\sigma') + \rho(\sigma')\,
\int d\cue\,\cue^{2}\, \phi^{\dagger}(\sigma,\cue) \phi(\sigma,\cue)
\nonumber\\
&-&\int d\cue \int d\cue'\,(\cue\cdot \cue')\,
\phi^{\dagger}(\sigma,\cue) \phi(\sigma,\cue)
\phi^{\dagger}(\sigma',\cue') \phi(\sigma',\cue')\Bigg)\nonumber\\
&+&\sum_{\sigma}\lambda(\sigma)\left(\int d\cue\,
 \phi^{\dagger}(\sigma,\cue) \phi(\sigma,\cue) -\rho(\sigma)\right),
\eq
where we have defined
\be
G(\sigma,\sigma')=\frac{\mathcal{E}(\sigma,\sigma')+
\mathcal{E}(\sigma',\sigma)}{|\sigma-\sigma'|}.
\ee

Next, we introduce the interaction term. Two kinds of interaction
vertices, corresponding to $\phi^{\dagger}$ creating a solid line
or $\phi$ destroying a solid line, are pictured in Fig.3.
\begin{figure}[t]
\centerline{\epsfig{file=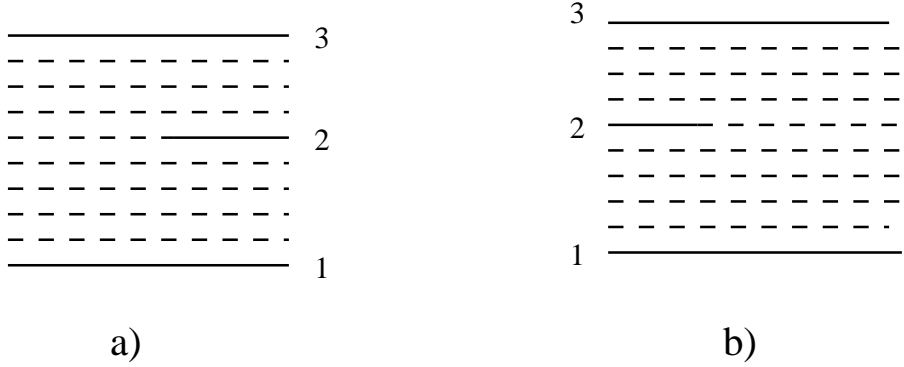, width=12cm}}
\caption{The Two $\phi^{3}$ Vertices}
\end{figure}
 We also
have to take care of the prefactor $1/(2 p^{+})$ in (1) by attaching it
to the vertices. Here, as in [4], we choose a symmeteric distribution of 
this factor between vertices by attaching a factor of
\be
V=\frac{1}{\sqrt{8\, p_{12}^{+}\, p_{23}^{+}\, p_{13}^{+}}}=
\frac{1}{\sqrt{8\,(\sigma_{2}-\sigma_{1})(\sigma_{3}-\sigma_{2}) 
(\sigma_{3}-\sigma_{1})}}
\ee
to each vertex. The interaction term in the hamiltonian can
now be written as
\be
H_{I}= g \sqrt{a}\,\sum_{\sigma}\int d\cue\,\left(\mathcal{V}(\sigma)\,
\rho_{+}(\sigma)\, \phi(\sigma,\cue)+
\rho_{-}(\sigma)\,\mathcal{V}(\sigma)\,
\phi^{\dagger}(\sigma,\cue)\right),
\ee
where $g$ is the coupling constant. $\rho_{\pm}$ are given by eq.(8) and
\be
\mathcal{V}(\sigma)=\dbsm \frac{W(\sigma_{1},\sigma_{2})}
{\sqsm},
\ee
where,
\be
\wss=\rho(\sigma_{1})\, \mathcal{E}(\sigma_{1},\sigma_{2})\,
 \rho(\sigma_{2}).
\ee

Here is a brief explanation of the origin of various terms in $H_{I}$:
The factors of $\rho_{\pm}$ are there to pair a solid line
with an $i=2$ fermion and a dotted line with an $i=1$ fermion. The
factor of $\mathcal{V}$ ensures that the pair of solid lines 12 and 23
in Fig.3
are seperated by only dotted lines, without any intervening solid lines.
Apart from an overall factor, the vertex defined above is very similar
 to the bosonic string interaction vertex in the light cone
picture. Taking advantage of the properties of $\mathcal{E}$ discussed
following eq.(5), we have written an explicit representation of 
this overlap vertex. Finally, the factor of $\sqrt{a}$ multiplying the
 coupling constant comes from   the replacement
$$
\int d\sigma \rightarrow a\,\sum_{\sigma}.
$$
It is easy to verify that the factor $a$ that appears on the right is
taken care of by attaching a factor of $\sqrt{a}$ to the coupling
constant g.

 The total hamiltonian is given by
\be
H=H_{0}+H_{I}
\ee
and the corresponding action by
\be
S=\int d\tau\left(\sum_{\sigma}\left(i \bar{\psi} \partial_{\tau}
\psi + i\int d\cue\,\phi^{\dagger} \partial_{\tau} \phi \right)
- H(\tau)\right).
\ee

For later use, we note that the theory is invariant under
\be
\phi(\sigma,\tau, \cue)\rightarrow \phi(\sigma,\tau, \cue+{\bf r}),
\ee
where ${\bf r}$ is a constant vector.

\vskip 9pt

\noindent{\bf 4. The Variational Treatment Of $\phi^{3}$ In 1+2 
Dimensions}

\vskip 9pt

In this section, we apply the variational ansatz introduced in [3] to
$\phi^{3}$ in 1+2 dimensions (D=1). As explained in the introduction,
in this low dimension there is no ultraviolet divergence, so we can focus
on the problem of removing the infrared cutoff by letting $a\rightarrow 0$
without being distracted by the ultraviolet problem.
An alternative and completely equivalent approach, developed in [4], is to 
search for a static classical
 solution to the action (16). Here we prefer the
 variational method since  it provides a rigorous upper bound to the ground
 state energy, which will be needed in the subsequent development. We choose
the variational state
\be
|s\rangle= \prod_{\sigma}\left(\int d\cue\,A(\cue)\,\phi^{\dagger}
(\sigma,\cue)\,\rho_{-}(\sigma)+ B\right)|0\rangle,
\ee
where the product extends over all $\sigma$ and the vacuum, which satisfies
$$
\rho_{+}(\sigma)|0\rangle =0,
$$
corresponds to the empty world sheet. By suitable gauge fixing (eq.(7)
and the discussion that follows),
A and B can be taken to be real, and they are $\sigma$ independent
to have a  ground state that is translationally invariant in $\sigma$.
Also, assuming a rotationally invariant ground state, A can only
depend on the length of the vector $\cue$.
 This trial state, 
introduced in [3], is just about simplest ansatz for the ground state
one can think of.

  Solving
  the constraint (3)
and the normalization condition gives
\be
\int d\cue\,A^{2}(\cue)=\rho_{0},\,\,\,B=\sqrt{1-\rho_{0}^{2}},
\ee
where $0\leq \rho_{0}\leq 1$
 is the (constant) ground state expectation value of $\rho$.

The expectation values of $H_{0}$ and $H_{I}$ (eqs.(9) and (12)) are
easily calculated:
\bq
\langle s|H_{0}|s\rangle &=& \sum_{\sigma'>\sigma} G(\sigma,\sigma')
\left(\frac{1}{2} m^{2}\,\rho_{0}^{2}+ \rho_{0}\,\int d\cue\,
\cue^{2}\,A^{2}(\cue)\right)\nonumber\\
&+&\lambda_{0}\left(\int d\cue\, A^{2}(\cue)-\rho_{0}\right),
\nonumber\\
\langle s|H_{I}|s\rangle &=& 2 g \sqrt{a}
\,B\,\sum_{\sigma}\mathcal{V}(\sigma)
\int d\cue\, A(\cue).
\eq
where $\lambda_{0}$ is the (constant) expectation value of the lagrange
multiplier $\lambda$.

We now define the ground state energy by
\be
E_{g}=\langle s|H_{0}+ H_{I}|s\rangle,
\ee
and minimize it with respect to the variational parameters $\lambda_{0}$,
$\rho_{0}$ and the function $A(\cue)$. The variational equation
\be
\frac{\delta E_{g}}{\delta A(\cue)}=0
\ee
determines A:
\be
A(\cue)=- g \sqrt{a}\,\sqrt{1-\rho_{0}}\,\frac{\mathcal{V}(\sigma)}
{ \lambda_{0}+ \rho_{0}\,\cue^{2}\,\sum_{\sigma'>\sigma} 
G(\sigma,\sigma')}.
\ee

Next, we evaluate various terms in this equation, setting
$\rho(\sigma)=\rho_{0}$. It is convenient to define
\be 
\sum_{\sigma'>\sigma} G(\sigma,\sigma')=\frac{1}{a}\,F(\rho_{0}),
\,\,\,\mathcal{V}(\sigma)=\frac{\rho_{0}^{2}}{a^{3/2}}\,Z(\rho_{0}),
\ee
where,
\bq
F(\rho_{0})&=&-\frac{\rho_{0}\,\ln(\rho_{0})}{1-\rho_{0}},
\nonumber\\
Z(\rho_{0})&=&\sum_{n_{1}=0}^{\infty}\sum_{n_{2}=0}^{\infty}
\frac{(1-\rho_{0})^{n_{1}+n_{2}}}{\sqrt{(n_{1}+1)\,
(n_{2}+1)\,(n_{1}+n_{2}+2)}}.
\eq
Eq.(23) can now be rewritten as
\be
A(\cue)=- g \,\rho_{0}^{2}\,\sqrt{1-\rho_{0}}\,\frac{Z(\rho_{0})}
{a \lambda_{0}+ F(\rho_{0})\,\cue^{2}}.
\ee

So far, D, the dimension of the transverse space, has been
arbitrary, but now, we specialize to $D=1$. The variational equation
\be
\frac{\partial E_{g}}{\partial \lambda_{0}}=0,
\ee
fixes $\lambda_{0}$:
\be
(a \lambda_{0})^{3}=\frac{\pi^{2}}{4}\,g^{4}\,(1-\rho_{0})^{2}
\,\rho_{0}^{6}\,\frac{Z^{4}(\rho_{0})}{F(\rho_{0})}.
\ee
we note that both $A(\cue)$ and $a\,\lambda_{0}$ are stay finite
as $a\rightarrow 0$.

It is convenient to seperate the contribution of the mass term to $E_{g}$:
\bq
E_{g}&=& E_{m}+ E_{0},\nonumber\\
E_{m}&=& \frac{p^{+}}{2 a^{2}}\,m^{2}\,\rho_{0}\,F(\rho_{0}),
\nonumber\\
E_{0}&=& -\frac{p^{+}}{3 a^{2}}\,\rho_{0}^{3}\,
\left(4 \pi\,g^{2}\,(1-\rho_{0})\,Z^{2}(\rho_{0})
\right)^{2/3}\,\left(F(\rho_{0})\right)^{- 1/3}.
\eq

It remains to minimize $E_{g}$ with respect to $\rho_{0}$.
Fortunately, without any detailed  analysis, we can
deduce  some qualitative general features of the
two terms,  which enables us to answer 
the important question of whether the minimum occurs at $\rho_{0}=0$
or at a non-vanishing $\rho_{0}$.

 We first note that both terms are bounded;
  $E_{m}$ is positive semi-definite, $E_{0}$ is negative semi-definite.
Next, we need their behaviour near the end points.
 At $\rho_{0}=1$, $E_{0}$ vanishes and
$E_{m}$ reaches a positive value, so $E_{g}$ is also positive at this
point. On the
other hand, as $\rho_{0}\rightarrow 0$, both go to zero, but $E_{m}$
vanishes faster:
\be
E_{m}\rightarrow - \rho_{0}^{2}\,\ln(\rho_{0}),\,\,\,
E_{0}\rightarrow \rho_{0}^{5/3}.
\ee
The second limit follows from
\be
Z(\rho_{0})\rightarrow \rho_{0}^{- 1/2},\,\,\,
F(\rho_{0})\rightarrow \rho_{0}\,\ln(\rho_{0}).
\ee

For small $\rho_{0}$,
since $E_{m}$ vanishes faster then $E_{0}$, $E_{0}$ wins over $E_{m}$,
and  $E_{g}$ is therefore negative. As $\rho_{0}$ increases towards one,
 it changes sign and becomes positive.
 This is sketched in Fig.4.
\begin{figure}[t]
\centerline{\epsfig {file=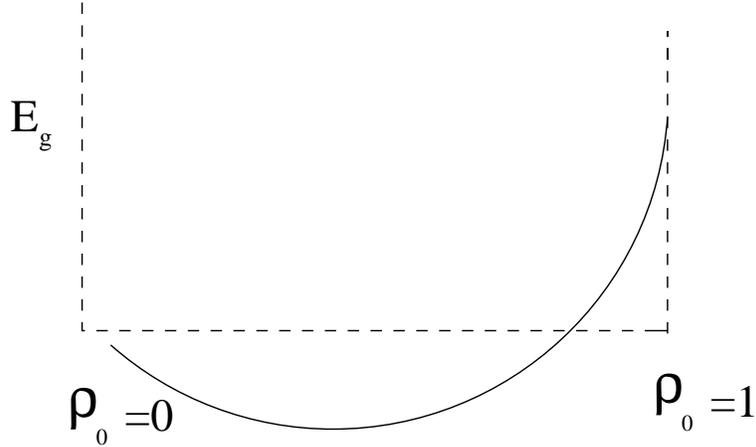, width=10cm}}
\caption{$E_{g}$ As  Function Of $\rho_{0}$}
\end{figure}
 Clearly, there is a
  minimum is at some $\rho_{0}\neq 0$, and $E_{g}$
is negative at the minimum.
 
 We recall that $\rho_{0}$,
the ground state expectation value of $\rho$, measures the average density
of graphs on the world sheet. It vanishes in any finite order
of perturbation theory,
 so it is natural to identify the phase 
 $\rho_{0}=0$ with the perturbative
regime. On the other hand, in the phase $\rho_{0}\neq 0$, the world sheet
is densely covered with graphs, and the contribution of higher (infinite)
order graphs dominate. We have seen above that it is this phase that is
 energetically favored.

In establishing the existence of a non-trivial ground state at a
$\rho_{0}\neq 0$, the negative sign of $E_{0}$ was crucial.
This is a subtle entropic effect related to the
counting of configurations. From the perspective of statistical mechanics,
$E_{g}$ is really the free energy
$$
F=E -T S
$$
which takes into account the entropy arising from the counting of the
world sheet graphs. Consider the state $|s\rangle$ of the variational
ansatz (eq.(18)). It represents a superposition of spin up states (dotted
lines) and spin down states (solid lines). At  $\rho_{0}=0$, the world sheet
is empty, corresponding to a single state with spin up at all sites. In 
this case, the entropy, and also the free energy is zero. On the other
hand, at $\rho_{0}\neq 0$, we have a superposition of a multitude of
spin up and spin down configurations, giving rise to non-zero entropy.
Clearly, it is the interaction term $H_{I}$ that generates the entropy
 by causing transitions
between spin up and down states. At large $g$,  $H_{I}$ dominates, the
 entropy increases and finally
drives $F$ to a negative value.

At this point, we face the problem of a negative
 ground state energy that
diverges as $1/a^{2}$ in the continuum limit $a\rightarrow 0$
 in the phase $\rho_{0}\neq 0$. We emphasize that this is not an
artifact of the approximation;
  the variational calculation
which puts an upper limit on the ground state energy shows that this is a
real effect. Of course, one obvious possibility is that,
since we are working with an intrinsically unstable field theory, we
should not be surprized to find a negative unbounded ground state energy.
However, in the next section, we show that
 the same phenomenon persists in an
at least classically stable theory, where, in addition to the original
$\phi^{3}$ term, the new model now has a positive $\phi^{4}$
interaction. Using the same variational wave function, we again
find a ground state at $\rho_{0}\neq 0$, with an energy that
goes as $- 1/a^{2}$.

 The problematic term appears to be a simple
additive term to the ground state energy. One can therefore simply
cancel this term by introducing a counter term in the action proportional
to the area of the world sheet. This is a familiar procedure in
string theory; however, a finite additive term remains undetermined.
This is fixed 
 by demanding Lorentz invariance in string theory [10]. Here,
unfortunately, Lorentz invariance is still an open problem.

\vskip 9pt

\noindent{\bf 5. The Variational Treatment Of $\phi^{3}+\phi^{4}$
 In 1+2 Dimensions}

\vskip 9pt

In this section, we add a $\phi^{4}$ interaction to the $\phi^{3}$
of the previous section. The idea is to have at least a classically
stable model. The total interaction hamiltonian is with this addition
is given by
\be
H_{I,t}= H_{I}+ H'_{I},
\ee
where $H_{I}$ is given by eq.(12) and,
\bq
H'_{I}&=& g'\,a\,\sum_{\sigma_{1},\sigma_{2},\sigma_{3},\sigma_{4}}\,
\int d \cue \int d \cue'\Bigg(
f_{1}\left(\sigma_{1},\sigma_{2},\sigma_{3},\sigma_{4}\right)
\,\phi^{\dagger}(\sigma_{2}, \cue)\, \phi(\sigma_{3}, \cue')\nonumber\\
&+& f_{2}\left(\sigma_{1},\sigma_{2},\sigma_{3},\sigma_{4}\right)
\left(\phi(\sigma_{2}, \cue)\,\phi(\sigma_{3},\cue') + \phi^{\dagger}
(\sigma_{2}, \cue)\,\phi^{\dagger}(\sigma_{3},\cue')\right)\Bigg).
\eq
Here, $g'$ is a positive coupling constant, scaled by $a$ in order
that in the limit $a\rightarrow 0$, sums over $\sigma$ smoothly go
over integrals over $\sigma$ (see section 3). The function $f_{1}$
is given by
$$
f_{1}\left(\sigma_{1},\sigma_{2},\sigma_{3},\sigma_{4}\right)=
\left((\sigma_{2} -\sigma_{1})\,(\sigma_{4} -\sigma_{2})\,
(\sigma_{4} -\sigma_{3})\,(\sigma_{3} -\sigma_{1})\right)^{- 1/2}
$$
for
$$
\sigma_{1}<\sigma_{2}<\sigma_{4},\,\,\,\sigma_{1}<\sigma_{3}<\sigma_{4},
$$
and it is zero otherwise. $f_{2}$ is given by
$$
f_{2}\left(\sigma_{1},\sigma_{2},\sigma_{3},\sigma_{4}\right)=
\left((\sigma_{2} -\sigma_{1})\,(\sigma_{4} -\sigma_{1})\,
(\sigma_{4} -\sigma_{3})\,(\sigma_{3} -\sigma_{2})\right)^{- 1/2}
$$
for 
$$
\sigma_{1}<\sigma_{2}<\sigma_{3}<\sigma_{4},
$$
and it is otherwise zero. $H'_{I}$ is the sum of three different four
point vertices pictured in Fig.5.
\begin{figure}
\centerline{\epsfig {file=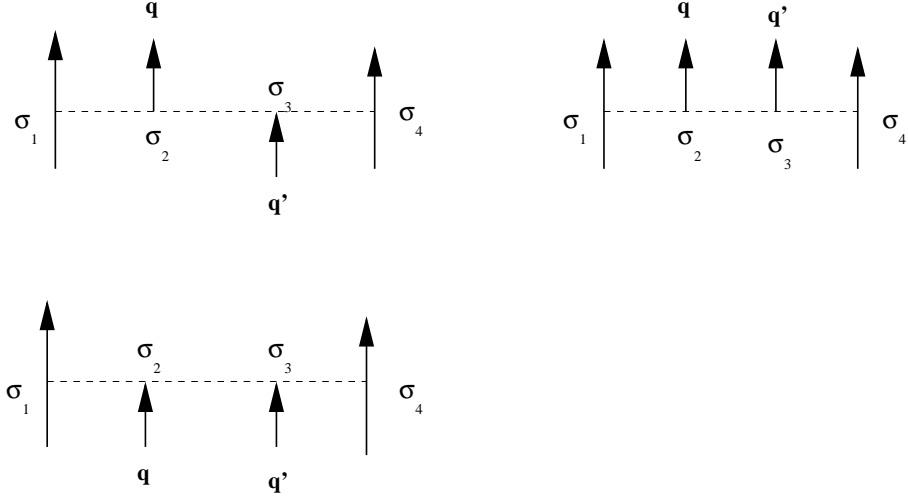, width= 12cm}}
\caption{Four Point Vertices}
\end{figure}

We will now do a variational calculation for this new model, using
the same form of the trial wave function given by eq.(18). The matrix
element of $H'_{I}$ is given by
\be
E'=\langle s|H'_{I} |s \rangle = \frac{p^{+}\,g'\,
\rho_{0}^{2}\,(1 -\rho_{0})\, \left( Z_{1}(\rho_{0})+ Z_{2}(\rho_{0})\right) }
{a^{2}}\,\int d \cue \int d \cue'\, A(\cue)\, A(\cue'),
\ee
where,
\bq
Z_{1}(\rho_{0})&=& \sum_{n_{1}=1}^\infty \sum_{n_{2}=1}^{\infty}
\sum_{n_{3}=1}^{\infty} \frac{(1- \rho_{0})^{n_{1}+n_{2}+n_{3}- 3}}
{\left(n_{1}\, n_{3}\,(n_{2}+ n_{3})\,(n_{1}+n_{2})\right)^{1/2}},
\nonumber\\
Z_{2}(\rho_{0})&=& \sum_{n_{1}=1}^\infty \sum_{n_{2}=1}^{\infty}
\sum_{n_{3}=1}^{\infty} \frac{(1- \rho_{0})^{n_{1}+n_{2}+n_{3}- 3}}
{\left(n_{1}\,n_{2}\,n_{3}\,(n_{1}+n_{2}+n_{3})\right)^{1/2}}.
\eq

Varying the ground state energy with the added term
\be
E_{g}= E_{m} +E_{0} + E'
\ee
with respect to $A(\cue)$ (see eq.(22)), we find the solution
\be
A(\cue)=\frac{h}{\pi}\,\frac{\left(a\,\lambda_{0}\,F(\rho_{0})\right)
^{1/2}}{a\,\lambda_{0}+\cue^{2}\,F(\rho_{0})},
\ee
and
\be
E'=\frac{p^{+}}{a^{2}}\,g'\,\rho_{0}^{2}\,(1-\rho_{0})\,\left(
Z_{1}(\rho_{0})+Z_{2}(\rho_{0})\right)\,h^{2},
\ee
where,
\be
h= \int d \cue\,A(\cue)= -\frac{g\,\pi\,\rho_{0}^{2}\,(1-\rho_{0})^{1/2}\,
Z(\rho_{0})}{\left(a\,\lambda_{0}\,F(\rho_{0})\right)
+2 \pi\,g'\,\rho_{0}^{2}\,
(1-\rho_{0})\left(Z_{1}(\rho_{0})+Z_{2}(\rho_{0})\right)}.
\ee
Finally, minimizing with respect to $\lambda_{0}$ (eq.(27)) gives the
relation
\be
\rho_{0}=\frac{h^{2}}{2 \pi}\,\left(\frac{F(\rho_{0})}{a\,\lambda_{0}}
\right).
\ee

Combining eqs.(39) and (40), one can solve for $\lambda_{0}$ and $h$,
 and express
$E'$  as a function of $\rho_{0}$ alone, and search for its minimum
in this variable. Here we will be satisfied with a qualitative analysis
similar to the one in the previous section. The new term in the ground state
energy, $E'$, is positive and vanishes at $\rho_{0}=0$. We also need its
 asymptotic behaviour as $\rho_{0}\rightarrow 0$. It is not difficult
to solve the above equations in this asymptotic limit. Starting with
$$
Z_{1,\, 2}(\rho_{0})\rightarrow 1/\rho_{0}, 
$$
and (31), we have,
$$
\lambda_{0}\rightarrow \rho_{0}\,\ln(\rho_{0}),\,\,\,
h\rightarrow \rho_{0}^{1/2},
$$
and
\be
E'\rightarrow \rho_{0}^{2}.
\ee

Therefore, for small enough $\rho_{0}$, the negative term $E_{0}$
wins over the positive terms $E_{m}$ and $E'$, and $E_{g}$
becomes negative. On the other hand, at $\rho_{0}=1$, $E_{0}$
vanishes, and $E_{g}$ becomes positive. The plot of $E_{g}$ against
$\rho_{0}$ again looks like Fig.4 , and we arrive at the same conclusion
as in the last section: The minimum of $E_{g}$ is at some
$\rho_{0}\neq 0$.

\vskip 9pt

\noindent{\bf 6.  Classical Solution And Renormalization In
1+3 Dimensions}

\vskip 9pt

One can easily apply the variational ansatz to the $\phi^{3}$ model
at $D=2$ (1+3 dimensions). However, a new complication arises: There
is a logarithmic mass divergence which has to be renormalized. The
standard recipe is to introduce a counter term in the bare mass to
cancel this divergence. The ground state energy, computed in
references [3, 4], is of the form
\be
E_{g}=\frac{p^{+}}{a^{2}}\,\left(- F_{1}(\rho_{0}) \,\ln\left(
F_{2}(\rho_{0})\,\Lambda^{2}\right) +\frac{1}{2}\,m^{2}\,\rho_{0}
\,F(\rho_{0})\right),
\ee
where $\Lambda$ is an ultraviolet cutoff,
$F_{1,2}$ and $F$ are certain functions of $\rho_{0}$ whose explicit 
form will not matter.
One can then cancel the cutoff dependent part of the energy
against the mass counter term $\delta m^{2}$ by letting
\be
m^{2}\rightarrow m_{r}^{2}+\delta m^{2}.
\ee

However, there is a problem with this;
 $\delta m^{2}$ will then be a function
of $\rho_{0}$, whereas in field theory the mass counter term can only
depend on the renormalized mass and the coupling constant.
Furthermore, if even if we decide to allow the counter term to depend
on $\rho_{0}$, this will inevitably introduce an arbitrary dependence
on this variable in $E_{g}$, and the position of the minimum can
no longer uniquely fixed in terms of the parameters of the original
field theory; namely, the mass and the coupling constant. We find this
situation unsatisfactory. 

Fortunately, we can modify our variational calculation to avoid this 
problem. First, we observe that the variational calculation is
equivalent to finding solutions to the classical equations of motion
[4]. For example, the classical equations motion with respect to $\phi$
and $\phi^{\dagger}$ can be solved for a static rotation invariant
configuration  in the form
\be
\phi_{0}(\sigma,\cue)= -g\,\sqrt{a}\,\frac{\rho_{-}(\sigma)
\,\mathcal{V}(\sigma)}{\lambda(\sigma)+
\frac{1}{2} G(\sigma,\sigma')\,\rho(\sigma')\,\cue^{2}}.
\ee
Replacing $\rho(\sigma)$ and $\rho_{-}(\sigma)$ by their expectation
values $\rho_{0}$ and $\sqrt{1 -\rho_{0}}$ respectively, 
we see that the classical solution
 $\phi_{0}$ is identical $A(\cue)$ (eq.(23)).  Therefore,
 the variational and the
classical equation approaches are equivalent.
 We chose the variational
approach in section 4 because it enabled us to put an upper bound 
on the ground state energy. 

Being equivalent to the variational calculation, the classical approximation
to the ground state energy has the same ultraviolent divergent term
at $D=2$ and
therefore suffers from the same problem. However, it provides a convenient
starting point for a modified approach which overcomes this problem.
It turns out that treating $\rho$ as a classical variable is at the root
of the problem. This implies factorization of the expectation value of
 products. For example,
\be
\langle \rho^{2} \rangle \rightarrow \langle \rho \rangle\,
 \langle \rho \rangle= \rho_{0}^{2},
\ee
and similarly for higher products. This is, by the way, also
equivalent to the mean field approximation.
 On the other hand, treated exactly, $\rho$
takes on only the disctrete values $0$ and $1$, and satisfies the
identities
\be
\rho^{2}(\sigma)=\rho(\sigma),\,\,\rho_{+}(\sigma)\rho_{-}(\sigma)= 
1-\rho(\sigma),\,\,\rho_{-}(\sigma)\rho_{+}(\sigma)=\rho(\sigma).
\ee
From these, one can derive two further identities
\bq
G(\sigma,\sigma')\, \rho(\sigma')\,\rho_{-}(\sigma)\,\wss &=&\nonumber\\
\left(\delta_{\sigma',\sigma_{2}}\,\frac{1}{\sigma_{2}- \sigma}+
\delta_{\sigma',\sigma_{1}}\,\frac{1}{\sigma -\sigma_{1}}\right)
&\rho_{-}(\sigma)&\wss,
\eq
and
\be
\wss \rho_{+}(\sigma) \rho_{-}(\sigma) W(\sigma'_{1},\sigma'_{2})
=\delta_{\sigma_{1},\sigma'_{1}}\,\delta_{\sigma_{2},\sigma'_{2}}   
\,\wss.
\ee

 One can also understand these geometrically from the overlap
properties of the vertices in Fig.3.  Apart
from an overall factor, these are structurally the same as the
corresponding string vertices, and in particular, they satisfy the
same overlap relations. These overlap relations turn out to be
crucial for preserving the correct structure of the ultraviolet
divergent terms. The factorization ansatz (45) violates the
identities (46) and consequently the above overlap relations.

This poses a dilemma: We cannot carry out an exact calculation all the
 way through and so we must eventually make some approximation. There is,
however, a way out: We carry out an exact calculation till we get
the structure of the ultraviolet divergence right, and only then we
resort to the classical (meanfield) approximation. The basic strategy
is first to simplify $\phi_{0}$ and later $E_{g}$ as much as possible
using the relations (46),(47) and (48) before making any approximations.
For example, $\phi_{0}$ can be rewritten as
\be
\phi_{0}(\sigma,\cue)=
 -g \dbsm\frac{\rho_{-}(\sigma)\,\wss}{\left(\lambda(\sigma)+
\frac{1}{2} \cue^{2}\left(\ssm\right)\right) \sqsm}.
\ee
This is
derived by formally expanding the denominator in eq.(44) in powers
of $\cue^{2}$ and then using the identity (47) repeatedly to simplify
products of the form $G\,W$. This is as far as one can go;
 since this equation is
linear in $W$, no further simplification is possible.

Next, we replace $\phi$ by the above $\phi_{0}$
 in the hamiltonian. The result involves
products of the form
$$
\rho_{+}\,\rho_{-},\,\,\,g\,W,\,\,\,   W\,W
$$
which can be simplified with the help of (46),(47) and (48), and the integral
over $\cue$ can be done. We skip the algebra and give the final result: 
\bq
H(\phi=\phi_{0})
&=&- 2\pi\,g^{2}\,a\,\sum_{\sigma}\dbsm \frac{\wss}{(\sigma_{2}
-\sigma_{1})^{2}}\,\ln\left(\frac{\Lambda^{2}}{\lambda(\sigma)}
\ssm\right)\nonumber\\
&+&\frac{m^{2}}{2}\,
\sum_{\sigma_{1}<\sigma_{2}} \frac{\wss}{|\sigma_{1}-\sigma_{2}|}
- \sum_{\sigma} \lambda(\sigma)\,\rho(\sigma),
\eq
where $\Lambda$ is an ultraviolet cutoff needed because of mass divergence.
Again, since this expression is linear in $W$, it cannot be
simplified any further.

The cutoff dependent part of the above hamiltonian exactly matches the
mass term in $H_{0}$ (eq.(9)). It can therefore be cancelled by letting
$$
m^{2}\rightarrow m_{r}^{2}+\delta m^{2},
$$
where $m_{r}$ is the renormalized mass and the counter term is
\bq
\delta m^{2}&=&
- 4\pi g^{2}\,a\,\sum_{\sigma}
 \dbsm \frac{\wss}{(\sigma_{2}-\sigma_{1})^{2}}
\,\ln\left(\Lambda^{2}/\mu^{2}\right)\nonumber\\
&=& - 4 \pi\,g^{2}\, \sum_{\sigma_{1}<\sigma_{2}}
\frac{\wss}{|\sigma_{1}-\sigma_{2}|}\,\ln\left(\Lambda^{2}/\mu^{2}\right).
\eq
 The renormalized  hamiltonian is then given by
\bq
H_{r}(\phi=\phi_{0})
&=&- 2\pi\,g^{2}\,a\,\sum_{\sigma}\dbsm \frac{\wss}{(\sigma_{2}
-\sigma_{1})^{2}}\,\ln\left(\frac{\mu^{2}}{\lambda(\sigma)}
\ssm\right)\nonumber\\
&+&\frac{m_{r}^{2}}{2}\,
\sum_{\sigma_{1}<\sigma_{2}} \frac{\wss}{|\sigma_{1}-\sigma_{2}|}
- \sum_{\sigma} \lambda(\sigma)\,\rho(\sigma),
\eq
In this result, the parameter $\mu$ is actually redundant; a change
in $\mu$ can always be absorbed into the definition of $m_{r}$. We
have therefore succeeded in renormalizing the mass divergence without
introducing any extra parameters not present in the original model.

With the renormalization done, we are ready to evaluate (52) in the classical
(mean field) approximation to compute the ground state energy.
 This amounts to replacing $\rho(\sigma)$ and
$\lambda(\sigma)$ by their $\sigma$ independent ground state
expectation values $\rho_{0}$ and $\lambda_{0}$. Furthermore, $\lambda_{0}$
can be fixed through the variational equation
$$
\frac{\delta H_{r}}{\delta \lambda_{0}}=0,
$$
with the result
\be
\lambda_{0}=\frac{2\,\pi\,g^{2}}{a}\,F(\rho_{0}).
\ee
Finally, putting everything together, the ground state energy is
\bq
E_{g}&=&H_{r}\left(\phi=\phi_{0},\,\rho=\rho_{0},\,\lambda=\lambda_{0}
\right)\nonumber\\
&=&\frac{p^{+}}{a^{2}}\Bigg(\left(\frac{1}{2} m_{r}^{2} - 6 \pi\,g^{2}\right)
\,\rho_{0}\,F(\rho_{0}) - 2\,\pi\,g^{2}\,\rho_{0}^{2}\,\tilde{F}(\rho_{0})
\nonumber\\
&-& 2\,\pi\,g^{2}\,\rho_{0}\,F(\rho_{0})\,\ln\left(\frac{\mu^{2}}{
2\,\pi\,g^{2}\,F(\rho_{0})}\right)\Bigg),
\eq
where $F$ is given by (31) and
$$
\tilde{F}(\rho_{0})=\sum_{n=1}^{\infty}\frac{\ln(n)}{n}\,(1-\rho_{0})^{n-1}.
$$

We are now ready to study the minimum of $E_{g}$ as a function
 of $\rho_{0}$ in the same qualitative fashion, as we have done earlier.
Firstly, since $\mu$ is redundant, the above expression can be
simplified without loss of generality by setting
$$
\mu^{2}=2\,\pi\,g^{2}.
$$
We then note that all the terms are bounded and they vanish at
$\rho_{0}=0$. Furthermore, they are all negative semi-definite with
the exception of the term proportional to $m^{2}$, which is positive
semi-definite. As before, we study the behaviour of various terms 
near the two end points $\rho_{0}= 0,\,1$. For sufficiently small
$\rho_{0}$, $\tilde{F}$ dominates over the positive mass term, so
$E_{g}$ is negative. Near $\rho_{0}=1$, if 
$$
m_{r}^{2}>12 \pi\,g^{2},
$$
$E_{g}$ is positive, and we are back to Fig.4 , with a minimum
at $\rho_{0}\neq 0$. If, on the other hand,
$$
m_{r}^{2}\leq 12 \pi\,g^{2}
$$
the situation is more complicated, although in this case also, the 
minimum is away from zero.

\vskip 9pt

\noindent{\bf 7. Classical Solution And Renormalization In 1+5
Dimensions}

\vskip 9pt

Next we consider $D=4$, corresponding to $\phi^{3}$ in six dimensions.
The self mass is now quadratically divergent, but this divergence can 
be eliminated by a mass counter term exactly as in the case $D=2$.
There remains, however, a residual logarithmic divergence:
\bq
H_{r}(\phi=\phi_{0})
&=& -4 \pi^{2}\,g^{2}\,a\,\sum_{\sigma}\dbsm \lambda(\sigma)
\,\wss \,\frac{(\sigma_{2}-\sigma) (\sigma -\sigma_{1})}
{(\sigma_{2} -\sigma_{1})^{3}}\nonumber\\
&\times& \ln\left(\frac{\lambda(\sigma)}{\Lambda^{2}} \ssm\right)
-\sum_{\sigma} \lambda(\sigma)\,\rho(\sigma)\nonumber\\
&+&\frac{m_{r}^{2}}{2}\,
\sum_{\sigma_{1}<\sigma_{2}} \frac{\wss}{|\sigma_{1}-\sigma_{2}|}.
\eq
This divergence can be eliminated by renormalizing the bare coupling
constant $g$ by setting
\be
g^{2}=\frac{g_{r}^{2}}{\ln\left(\Lambda^{2}/\mu^{2}\right)},
\ee
where $g_{r}$ is the renormalized coupling constant and $\mu$ an
arbitrary mass parameter. We recall that $\phi^{3}$ is asymptotically free
in 6 space-time dimensions, and the above relation between the bare and
renormalized couplings is the well known lowest order result.
 In the limit $\Lambda\rightarrow \infty$,
the renormalized $H_{r}$ is given by
\bq
H_{r}(\phi=\phi_{0})&\rightarrow& \sum_{\sigma}\lambda(\sigma)
\left( 4 \pi^{2} g_{r}^{2}\,a \dbsm 
\wss \,\frac{(\sigma_{2}-\sigma) (\sigma -\sigma_{1})}
{(\sigma_{2} -\sigma_{1})^{3}} - \rho(\sigma)\right)\nonumber\\
&+&\frac{m_{r}^{2}}{2}\,
\sum_{\sigma_{1}<\sigma_{2}} \frac{\wss}{|\sigma_{1}-\sigma_{2}|}.
\eq

Having renormalized the ultraviolet divergences, we ready to 
carry out the mean field (classical) approximation. As before, we
replace $\rho$ by its constant expectation value $\rho_{0}$.
But in the case of $\lambda$, we encounter a new situation, different
from the dimensions $D=1,\,3$. h
The expectation value of $\lambda$ is undetermined; instead,
$\lambda$ acts as a lagrange multiplier and imposes the constraint
\bq
\rho_{0}&=& 4 \pi^{2}\,g_{r}^{2}\,a\,\dbsm
\left( \wss \,\frac{(\sigma_{2}-\sigma) (\sigma -\sigma_{1})}
{(\sigma_{2} -\sigma_{1})^{3}} \right)_{\rho=\rho_{0}}\nonumber\\
&=& 4 \pi^{2}\,g_{r}^{2}\,\rho_{0}^{2}\,\sum_{n_{1}=1}^{\infty} 
\sum_{n_{2}=1}^{\infty} \frac{n_{1}\,n_{2}}{(n_{1}+n_{2})^{3}}\,
(1 -\rho_{0})^{n_{1}+n_{2} -1}, 
\eq
and the ground state energy reduces to the mass term:
\be
E_{g}=H_{r}(\phi=\phi_{0},\,\rho=\rho_{0})=
\frac{p^{+}}{2 a^{2}}\,m^{2}_{r}\,\rho_{0}\,F(\rho_{0}).
\ee

Eq.(58) is the equation that fixes $\rho_{0}$. It has two solutions:
The trivial (perturbative ) solution
$$
\rho_{0}=0
$$
minimizes the ground state energy at
$$
E_{g}=0
$$
if $m_{r}^{2}>0$. On the other hand, for some range of $g_{r}^{2}$,
there is also the non-perturbative solution at some
$$
\rho_{0}\neq 0.
$$
This the solution that minimizes the energy if $m_{r}^{2}<0$. Normally,
one would tend to reject this possibility because of the tachyonic
mass. However, in the case of the asymptotically free theory that we
are dealing with, $m_{r}$ is likely to be a mass scale associated
with the running coupling constant, unrelated to the low energy
physical mass spectrum At this point, one should keep an open mind,
and a deeper study of the model is needed to decide on the sign
of $m_{r}^{2}$.

\vskip 9pt

\noindent{\bf 8. Quadratic Fluctuations Around The Classical
Background} 

\vskip 9pt

This section is mostly based on the earlier work on the same problem,
 especially reference [4]. There will, however, be some additions,
 corrections and clarifications. Our goal is to determine the spectrum
of quadratic fluctuations around the classical solution $\phi_{0}$ in the
phase $\rho_{0}\neq 0$. We are especially interested in the 
decompactification limit $a\rightarrow 0$: In this limit,
 the spectrum of 
fluctuations consists of a heavy sector and a light sector, and
 the masses of the heavy sector
excitations go to infinity, whereas the light sector masses remain finite.
It is natural to identify the heavy sector with the Kaluza-Klein
modes generated by the compactification; they become infinitely heavy
in the  decompactification limit
$$
a\rightarrow 0,\,\,\,R=1/a\rightarrow \infty.
$$
 Interestingly, the spectrum of the
light sector is that of a transverse string. This is in agreement with
the resuts of [4].

We now sketch a brief derivation of these results.
It is convenient to set
\be
\phi=\phi_{0}+\phi_{1},\,\,\,
\phi_{1}= \phi_{1,r}+i\, \phi_{1,i},
\ee
where
$$
\phi_{0}= A(\cue)
$$
 for $d=1$, and $\phi_{1,r,i}$ are hermitian fields. The contribution
 to the action second order in $\phi_{1}$ is given by
the sum of kinetic and potential terms:
\be
S^{(2)}= S_{k.e}- \int d\tau\,H^{(2)}(\tau)=S_{k.e}+ S_{p.e},
\ee
where,
\be
S_{k.e}=2 \sum_{\sigma} \int d\tau \int d\cue\,\phi_{1,i}\,
\partial_{\tau}\phi_{1,r},
\ee
 Since the action is quadratic in both $\phi_{1,i}$ or $\phi_{1,r}$,
 one can carry out the 
functional integral over one of these fields before writing down $H^{(2)}$.
 We choose to integrate over
 $\phi_{1,i}$, with the result,
\be
S_{k.e}\rightarrow  \sum_{\sigma} \int d\tau \int d\cue\,
\frac{\left(\partial_{\tau}\phi_{1,r}(\sigma,\tau,\cue)\right)^{2}}
{\lambda(\sigma)+ \frac{1}{2} \sum_{\sigma'} G(\sigma,\sigma')\,
\rho(\sigma')\,\cue^{2}},
\ee
and, somewhat schematically,
\bq
H^{(2)}&\rightarrow&\sum_{\sigma} \lambda(\sigma) \int d\cue\,
\phi_{1,r}^{2}(\sigma,\cue)+
\sum_{\sigma,\sigma'} G(\sigma,\sigma') \Bigg(\frac{1}{2}\,
\rho(\sigma')\,\int d\cue\,
\cue^{2}\,\phi_{1,r}^{2}(\sigma,\cue)\nonumber\\
& -& 2 \int d\cue \int d\cue' (\cue\cdot \cue')\,
(\phi_{0}\,\phi_{1,r})_{\sigma,\cue}\,(\phi_{0}\,\phi_{1,r})_{\sigma',
\cue'}\Bigg).
\eq

 We first fix $\rho$ and $\lambda$ at their classical values
$\rho_{0}$ and $\lambda_{0}$. Later, we will also consider their
fluctuations around the classical values. The first 
observation is that $S_{k.e}$ reaches a finite limit as 
 $a\rightarrow 0$. To see this, in this limit, it can be rewritten
as
\be
S_{k.e}\rightarrow  \int d \sigma \int d\tau \int d\cue\,
\frac{\left(\partial_{\tau}\phi_{1,r}(\sigma,\tau,\cue)\right)^{2}}
{a\,\lambda_{0}+ \frac{a}{2} \sum_{\sigma'} G(\sigma,\sigma')\,
\rho_{0}\,\cue^{2}},
\ee
and since $(a\,\lambda_{0})$ as well as
$$
a \sum_{\sigma'} G(\sigma,\sigma')
$$
are finite as  $a\rightarrow 0$ (see eq.(24)), so is $S_{k.e}$.
On the other hand, to leading order
$$
H^{(2)}\rightarrow 1/a^{2},
$$
and
therefore, in general, the spectrum becomes heavy in the decompactification
limit. This argument, with a slight modification, also applies to
the fluctuations of $\rho$ and $\lambda$. These fluctuations
become heavy and  are
suppressed in the limit $a\rightarrow 0$. Consequently, in this limit,
these fields become frozen at their expectation values
$\rho_{0}$ and $\lambda_{0}$.

 There are, however, exceptional modes which stay light.
To investigate $a\rightarrow 0$ limit more carefully, following [7],
 we note that the term involving  $G$ in $H^{(2)}$ can be written as
$$
\frac{1}{2} \sum_{\sigma, \sigma'} G(\sigma, \sigma')\,
K(\sigma)\,L(\sigma').
$$
Next, we expand the product $K L$ in powers of $\sigma' -\sigma$:
\be
K(\sigma) \,L(\sigma')=K(\sigma) \,L(\sigma)+(\sigma' -\sigma)\,
K(\sigma) \,L'(\sigma)+ \frac{1}{2} (\sigma' -\sigma)^{2}\,
K(\sigma) \,L''(\sigma)+\cdots,
\ee
and evaluate the sums over $\sigma'$:
\bq
&&\sum_{\sigma'} G(\sigma, \sigma')= \frac{2}{a}\,F(\rho_{0}),\nonumber\\
&&\sum_{\sigma'} (\sigma' -\sigma)\, G(\sigma, \sigma')=0,\nonumber\\
&&\sum_{\sigma'} (\sigma' -\sigma)^{2}\, G(\sigma, \sigma')
=\frac{2 a}{\rho_{0}^{2}}.
\eq
With the help of this expansion, we have, as $a\rightarrow 0$,
\bq
&&\frac{1}{2} \sum_{\sigma, \sigma'} G(\sigma, \sigma')\,
K(\sigma)\,L(\sigma')\rightarrow \nonumber\\
&&\int d\sigma \left(\frac{1}{a^{2}}\,F(\rho_{0})\,K(\sigma)\,L(\sigma)
-\frac{1}{2 \rho_{0}^{2}}\,K'(\sigma)\,L'(\sigma)\right).
\eq
If we had continued the expansion in eq.(66) to higher powers of
$\sigma' -\sigma$, we would have additional terms in the above expression
involving higher derivatives of $K$ and $L$, but their coefficients would
contain powers of $a$ and would therefore vanish.

It is of interest to note that although we started with an action
non-local in the $\sigma$ coordinate, after taking the $a\rightarrow 0$
limit, we end up with a local action, with at most two derivatives
with respect to $\sigma$. This localization is a concequence of
$\rho_{0}$ being non-zero. To see this, consider  the term in
$H_{0}$ involving the product 
$$
\left(\phi^{\dagger} \phi\right)_{\sigma}\,
\left(\phi^{\dagger} \phi\right)_{\sigma'}.
$$
The points $\sigma$ and $\sigma'$ are tied together by a factor of 
$\mathcal{E}(\sigma, \sigma')$ (eqs.(6) and (14)), which, after
setting $\rho(\sigma)=\rho_{0}$, becomes
$$
\mathcal{E}\rightarrow (1- \rho_{0})^{n -1},
$$
where,
$$
n=\frac{|\sigma -\sigma'|}{a}.
$$
If we keep $\sigma -\sigma'$ fixed and finite
 as we let $a\rightarrow 0$,
$n\rightarrow \infty$, and since $(1- \rho_{0})<1$,
$$
\mathcal{E}(\sigma, \sigma')\rightarrow 0.
$$
Therefore, in the sum over $\sigma$ and $\sigma'$ in (68), only
the terms seperated by a distance of the order of $a$ survive,
and the theory localizes.

It was noted in the earlier work [7] that there is an exceptional mode
for which the leading term in $H^{(2)}$ proportional to $1/a^{2}$
is absent, and therefore  this mode survives in the
decompactification limit. This is not an accident but it is the result
of the invariance of the theory under
\be
\cue\rightarrow \cue +{\bf r},
\ee
(see eq.(17)). Now consider the mode $\bfv$
introduced by letting
\be
\phi_{1,r}\rightarrow \phi_{0}(\sigma,\,\cue+\bfv(\sigma, \tau))
- \phi_{0}(\sigma, \cue),
\ee
where $\phi_{0}$ is the classical solution. It is easy to see that 
the leading contribution of $\bfv$  to $S_{p.e}$
 would be a mass term for proportional to
$$
\frac{1}{a^{2}}\,\int d\tau \int d\sigma \,\bfv^{2}(\sigma,\tau).
$$
But such a term is forbidden by the translation symmetry, which can
be rephrased as a symmetry under
\be
\bfv(\sigma,\tau)\rightarrow \bfv(\sigma,\tau)+ {\bf r}.
\ee
So only the finite term $K' L'$ in (68) survives.
This is a familiar story: $\bfv$ can be identified with the Goldstone
mode of spontaneously broken translation invariance.

The contribution of $\bfv$ to $S_{p.e}$ is easily calculated
by making the substitution (70) in (64). The integral over $\cue$
can be done by simply shifting
$$
\cue \rightarrow \cue+ \bfv,
$$
with the result
\be
S_{p.e}(\bfv)=-\frac{1}{2} \int d\tau \int d\sigma\,
\left(\partial_{\sigma}\bfv\right)^{2}.
\ee
This result is universal in the sense that it does not depend
on the structure of $\phi_{0}$ or the dimension $D$. On the other hand,
the kinetic energy term does depend on both $\phi_{0}$ and $D$.
Making the substitution (70) in (63), after a short calculation, we have,
\be
S_{k.e}(\bfv)=5\,\pi \, g^{2} (4 \,a\,\lambda_{0})^{- 7/2}
\rho_{0}^{2}\,(1-\rho_{0})\,F_{1}(\rho_{0})
 \int d\tau \int d\sigma\,
\left(\partial_{\tau}\bfv\right)^{2},
\ee
where $a\,\lambda_{0}$ is given by (28) and
$$
F_{1}(\rho_{0})=\sum_{n_{1}=1}^{\infty} \sum_{n_{2}=1}^{\infty}
\frac{(1 -\rho_{0})^{n_{1}+n_{2}-2}}{n_{1}^{3/2}\,n_{2}^{3/2}\,
(n_{1}+n_{2})^{1/2}}.
$$
Combining (72) and (73),
\be
S^{(2)}=\int d\tau \int_{0}^{p^{+}} d\sigma \left(8 \pi^{2}\,
\alpha'^{2}\,\left(\partial_{\tau}\bfv\right)^{2}
- \frac{1}{2}\,\left(\partial_{\sigma}\bfv\right)^{2}\right).
\ee

This is the action for a transverse string, where the slope is given by
\be
\alpha'^{2}=\frac{5\,g^{2}\,\rho_{0}^{2}\,(1-\rho_{0})\,
F_{1}(\rho_{0})}{8 \pi}
\,\left(4\,a\,\lambda_{0}\right)^{- 7/2}.
\ee
The detailed form of this expression is not important; what is important
is that the slope is non-zero for $\rho_{0}\neq 0$. A similar
calculation for $D=2$ also gives the same result: The slope is 
non-zero for non-zero $\rho_{0}$.

The situation in the case $D=4$ is  problematic. In addition to
demanding that $\rho_{0}$ be non-zero, which requires a tachyonic mass term
(see the discussion in section 7), we have also to specify that
$(a\,\lambda_{0})$ is finite. We recall that unlike in the lower
dimensional cases, at $D=4$, $\lambda_{0}$ is undetermined.
So in this case, string formation remains an open problem.

\vskip 9pt

\noindent{\bf 9. Systematic Expansion}

\vskip 9pt

So far what we have done is to construct 
static classical solutions  and 
expand around them. It would be very desirable to identify this
expansion with a systematic expansion in powers of a free parameter.
In the case of $D=1$, we are able to do this. We remind the reader of
 the usual weak coupling expansion around a soliton. One first splits
the hamiltonian into the kinetic and potential terms:
\be
H= H_{V} + H_{K}.
\ee
After a a suitable scaling of the fields,  the hamiltonian is cast into the
form
\be
H\rightarrow \frac{1}{e^{2}}\,\tilde{H}_{V} +e^{\beta}\,\tilde{H}_{K}
\ee
where $e$ is the expansion parameter, $\beta$ is a positive number,
and $\tilde{H}_{V}$ and $\tilde{H}_{K}$, expressed in terms of scaled
fields, are independent of $e$. In the small $e$ limit,
the saddle point equations
$$
\frac{\partial\tilde{H}_{V}}{\partial\tilde{\phi}_{i}}=0
$$
determine the solitonic field configuration. One then expands in powers of
$e$ around this configuration, which becomes heavy in the weak coupling
limit. The kinetic term $\tilde{H}_{K}$ is also
treated as a perturbation.

In the case at hand, $D=1$, $H_{V}$ is the total hamiltonian (eq.(15)),
and $H_{K}$ is
\be
H_{K}=\sum_{\sigma}\left(i \bar{\psi} \partial_{\tau}
\psi + i\int d\cue\,\phi^{\dagger} \partial_{\tau} \phi \right).
\ee
We now define the tilde fields by
\bq
\phi(\sigma, \cue)&=& g^{- 1/3} \,\tilde{\phi}(\sigma, g^{-2/3}\,
\cue),\,\,\,
\phi^{\dagger}(\sigma, \cue)= g^{- 1/3} \,
\tilde{\phi}^{\dagger}(\sigma, g^{-2/3}\,\cue),\nonumber\\
\lambda(\sigma)&=& g^{4/3}\,\tilde{\lambda}(\sigma),
\eq
and also redefine the mass by
\be
m= g^{2/3}\,\tilde{m}.
\ee
The fermionic fields and $\rho$ are unchanged. Written in terms of these
new fields and ${\tilde m}$, $H_{V}$ scales as
\be
H_{V}(\phi, \lambda, m)= g^{4/3}\,
 H_{V}(\tilde{\phi}, \tilde{\lambda}, \tilde{m}).
\ee
The tilde variables are so chosen that when $H_{V}$ is expressed in
terms of them, it no longer depends on $g$, except for the overall 
factor of  $g^{4/3}$. We can therefore identify the expansion
parameter as
\be
e^{2}= g^{-4/3}.
\ee
and the expansion is a strong coupling expansion in inverse 
powers of $g^{4/3}$. As expected, the soliton becomes
heavy in this limit. This is a surprise, since in field theory,
solitons are usually associated with the weak coupling limit.

 It is easily verified that the classical
solution (26) transforms as $\phi$ does, and since
\be
E_{g}\rightarrow g^{4/3}\, E_{g},
\ee
$\rho_{0}$, the location of the
minimum of $E_{g}(\rho_{0})$, is unchanged. Also, the transformation
(79) preserves the canonical commutation relations (2). Finally, $S_{K}$
transforms with no prefactor ($\beta=0$ in (77)), and should therefore be 
treated as a perturbation. We also note that the string slope goes as
$$
\alpha'^{2}\rightarrow g^{- 8/3}
$$
for large $g$. The strong coupling limit is therefore the zero
slope limit, when only the low lying states survive. The resulting
model is then the strong coupling dual of the original model.

A similar strong coupling expansion should also be applicable 
$D=2$ case once the mass renormalization is taken care of.
 The case $D=4$ has additional complications
due to the running coupling constant and remains a completely
open problem.

\vskip 9pt

\noindent{\bf 10. Conclusions}

\vskip 9pt

The main new contribution of the present work are the items discussed
at the end of the introduction. They are the introduction of a 
$\phi^{4}$ interaction in addition to the original $\phi^{3}$,
the correct treatment of the ultraviolet divergences, and in $1+2$
dimensions, the development of a systematic 
strong coupling expansion around the
classical configuration
 in section 9. Since more realistic theories such as
gauge theories have $\phi^{4}$ type interactions, generalizations
of what was done here may lead towards more realistic models\footnote{
For some initial steps taken towards more realistic theories,
see [11, 12].}.
Also, a systematic expansion is always good news; it should be useful
in investigating problems like Lorentz invariance\footnote{See [13] for
an investigation of renormalization and Lorentz invariance in the
light cone formulation.} in a systematic fashion.

Many open problems still remain for future research. It would very
nice to extend the strong coupling expansion to dimensions
$1+3$ and $1+5$. Also, generalizations to at least lower dimensional
gauge theories appears to be within reach. And of course, the problem
of Lorentz invariance mentioned above still remains an important goal.

\vskip 9pt

\noindent{\bf Acknowledgement}

\vskip 9pt

This work was supported in part by the director, Office of Science,
Office of High Energy Physics of the U.S. Department of Energy under Contract
DE-AC02--05CH11231.

\vskip 9pt

\noindent{\bf References}

\vskip 9pt

\begin{enumerate}

\item K.Bardakci and C.B.Thorn, Nucl.Phys. {\bf B 626} (2002)
287, hep-th/0110301.
\item G.'t Hooft, Nucl.Phys. {\bf B 72} (1974) 461.
\item K.Bardakci, JHEP {\bf 0810} (2008) 056, arXiv:0808.2959. 
\item K.Bardakci, JHEP {\bf 0903} (2009) 088, arXiv:0901.0949.
\item H.P.Nielsen and P.Olesen, Phys.Lett. {\bf B 32} (1970) 203.
\item B.Sakita and M.A.Virasoro, Phys.Rev.Lett. {\bf 24} (1970) 1146.
\item K.Bardakci, JHEP {\bf 1003} (2010) 107, arXiv:0912.1304.
\item A.Casher, Phys.Rev. {\bf D 14} (1976) 452.
\item R.Giles and C.B.Thorn, Phys.Rev. {\bf D 16} (1977) 366.
\item P.Goddard, J.Goldstone, C.Rebbi and C.B.Thorn, Nucl.Phys. {\bf B 56}
(1973) 109.
\item C.B.Thorn, Nucl.Phys. {\bf B 637} (2002) 272, hep-th/0203167,
S.Gudmundsson, C.B.Thorn and T.A.Tran, Nucl.Phys. {\bf B 649} 92003)
3-38, hep-th/0209102.
\item C.B.Thorn and T.A.Tran, Nucl.Phys. {\bf B677} (2004) 289,
hep-th/0307203.
\item C.B.Thorn, Nucl.Phys. {\bf B 699} 427, hep-th/0405018,
D.Chakrabarti, J.Qiu and C.B.Thorn, Phys.Rev. {\bf D 74} (2006)
045018, hep-th/0602026.

\end{enumerate}

\end{document}